\documentclass[preprints,article,accept,moreauthors,pdftex]{mdpi}

\usepackage[american]{babel}
\usepackage[utf8]{inputenc}

\firstpage{1}
\pubvolume{xx}
\issuenum{1}
\articlenumber{5}
\pubyear{2020}
\copyrightyear{2020}
\history{}

\usepackage{amsmath}

\usepackage{url}
\def\UrlBreaks{\do\/\do-}

\Title{Accurate Sampling with Noisy Forces from Approximate Computing}

\abstract{
In scientific computing, the acceleration of atomistic computer simulations by means of custom hardware is finding ever growing application.
A major limitation, however, is that the high efficiency in terms of performance and low power consumption entails the massive usage of low-precision computing units. Here, based on the approximate computing paradigm, we present an algorithmic method to rigorously compensate for numerical inaccuracies due to low-accuracy arithmetic operations, yet still obtaining exact expectation values using a properly modified Langevin-type equation.
}


\Author{Varadarajan Rengaraj $^{1,3,\ddag}$, Michael Lass $^{3,4,\ddag}\orcidC{}$, Christian Plessl $^{3,4}$\orcidA{}, and Thomas D. Kühne $^{1,2}*$\orcidB{}}

\AuthorNames{Varadarajan Rengaraj, Michael Lass, Christian Plessl and Thomas D. Kühne}

\address{%
$^{1}$ \quad Dynamics of Condensed Matter, Department of Chemistry, Paderborn University, Warburger Str. 100, 33098 Paderborn, Germany ;\\
$^{2}$ \quad Center for Sustainable Systems Design, Paderborn University, Warburger Str. 100, 33098 Paderborn, Germany ;\\
$^{3}$ \quad Department of Computer Science, Paderborn University, Warburger Str. 100, 33098 Paderborn, Germany; \\
$^{4}$ \quad Paderborn Center for Parallel Computing, Paderborn University, Warburger Str. 100, 33098 Paderborn, Germany;
}

\corres{Correspondence: tdkuehne@mail.upb.de}

\secondnote{These authors contributed equally to this work.}

\begin{document}

\section{Introduction}

Molecular dynamics (MD) is a very powerful and widely used technique to study thermodynamic equilibrium properties, as well as the real-time dynamics of complex systems made up of interacting atoms \cite{AlderWainwright1957}. This is done by numerically solving Newton's equations of motion in a time-discretized fashion via computing the nuclear forces of all atoms at every time step \cite{RahmanMD}. Computing these forces by analytically differentiating the interatomic potential with respect to the nuclear coordinates is computationally rather expensive, which is particularly true for electronic structure based \textit{ab-initio} MD simulations \cite{CPMD, CPMD_TDK, PayneRMP, WIRES_TDK}.

For a long time newly developed microchips became faster and more efficient over time due to new manufacturing processes and shrinking transistor sizes. However, this development slowly comes to an end as scaling down the structures of silicon based chips becomes more and more difficult. The focus therefore shifts towards making efficient use of the available technology. Hence, beside algorithmic developments \cite{MTS, Snir, GSE, Shaw, VerletCell, pSHAKE, John, Prodan}, there have been numerous custom computing efforts in this area to increase the efficiency of MD simulations by means of hardware acceleration, which we take up in this work. Examples of the latter are MD implementations on graphics processing units (GPUs) \cite{HOOMD, NAMD, OpenMM, HalMD, Lammps, Amber, Gromacs}, field-programmable gate arrays (FPGAs) \cite{HerbordtI, HerbordtII}, and application-specific integrated circuits (ASICs) \cite{AntonI, AntonII}.
While the use of GPUs for scientific applications is relatively widespread \cite{GPUcomp,Binder,Weigel}, the use of ASICs \cite{QCDScience, QCDOC, GrapeScience, Grape} and FPGAs is less common \cite{JanusI, JanusII, Convey, FDTD, Kenter, Galerkin}, but gained attention over the last years.
In general, to maximize the computational power for a given silicon area, or equivalently minimize the power-consumption per arithmetic operation, more and more computing units are replaced with lower-precision units. This trend is mostly driven by market considerations of the gaming and artificial intelligence industries, which are the target customers of hardware accelerators and naturally do not absolutely rely on full computing accuracy.

In the approach presented in this paper, we mimic in software how it is possible to make effective use of low-accuracy special-purpose hardware for general-purpose scientific computing by leveraging the approximate computing (AC) paradigm~\cite{KlavikMalossiBekasEtAl2014, PlesslAC}. The general research goal of AC is to devise and explore ingenious techniques to relax the exactness of a calculation to facilitate the design of more powerful and/or more efficient computer systems. However, in scientific computing, where the exactness of all computed results is of paramount importance, attenuating accuracy requirements is not an option. Yet, assuming that the inaccuracies within the nuclear forces due to the usage of low-precision arithmetic operations can be approximately considered as white, we will demonstrate that it is nevertheless possible to rigorously compensate for such numerical errors and still obtain exact expectation values, as obtained by ensemble averages of a properly modified Langevin equation.

The remainder of the paper is organized as follows. In Section~\ref{sec:ac} we revisit the basic principles of AC before introducing our modified Langevin equation in Section~\ref{sec:methodology}. Thereafter, in Section~\ref{sec:computational}, we describe the computational details of our computational experiments. Our results are presented and discussed in Section~\ref{sec:results} before concluding the paper in Section~\ref{sec:conclusion}.

\section{Approximate Computing}
\label{sec:ac}

A basic method of approximation and a key requirement for efficient use of processing hardware is the use of adequate data widths in computationally intensive kernels. While in many scientific applications the use of double-precision floating-point is most common, this precision is not always required.
For example, iterative methods can exhibit resilience against low precision arithmetic as has been shown for the computation of inverse matrix roots~\cite{lass17-esl} and for solving systems of linear equations~\cite{KlavikMalossiBekasEtAl2014,Bekas,Dongarra2017,Dongarra2018}.
Mainly driven by the growing popularity of artificial neural networks \cite{Gupta2015}, we can observe growing support of low-precision data types
in hardware accelerators.
In fact, recent GPUs targeting the data center have started supporting half-precision as well, nearly doubling the peak performance compared to single-precision and quadrupling it compared to double-precision arithmetics~\cite{tesla}. However, due to the low number of exponent bits, half-precision only provides a very limited dynamic range. In contrast, \texttt{bfloat16} provides the same dynamic range as single-precision, and just reduces precision. It is currently supported by Google's Tensor Processing Units (TPU)~\cite{tpu} and support is announced for future Intel Xeon processors~\cite{xeon} and Intel AgileX FPGAs. A list of commonly used data types, together with the corresponding number of bits used to store the exponent and the mantissa, are shown in Table~\ref{tab:float} beside the double-precision \emph{de facto} standard. 

\begin{table}
  \caption{Bitwidth of common floating-point formats}
  \centering
  \label{tab:float}
  \begin{tabular}{lrrr}
    Type & sign & exponent & mantissa \\
    \hline
    IEEE 754 Quadruple-precision & 1 & 15 & 112 \\
    IEEE 754 Double-precision & 1 & 11 & 52 \\
    IEEE 754 Single-precision & 1 & 8 & 23 \\
    IEEE 754 Half-precision & 1 & 5 & 10 \\
    Bfloat16 (truncated IEEE single-precision) & 1 & 8 & 7
  \end{tabular}
\end{table}

Yet, programmable hardware such as FPGAs, as a platform for custom-built accelerator designs \cite{Strzodka2006, KenterVector, KenterPragma}, can make effective use of all of these, but also entirely custom number formats.
Developers can specify the number of exponent and mantissa bits and trade off precision against the amount of memory blocks required to store values and the number of logic elements required to perform arithmetic operations on them.

In addition to floating-point formats, also fixed-point representations can be used. Here, all numbers are stored as integers of fixed size with a
predefined scaling factor. Calculations are thereby performed using integer arithmetic. On CPUs and GPUs only certain models can perform integer operations with a peak performance similar to that of floating-point arithmetic, depending on the capabilities of the vector units / stream processors. Nevertheless, FPGAs typically can perform integer operations with performance similar to or even higher than that of floating-point. Due to the high flexibility of FPGAs with respect to different data formats and the possible use of entirely custom data types, we see them as the main target technology for our work. For this reason, we consider both floating-point and fixed-point arithmetic in the following.

\section{Methodology}
\label{sec:methodology}
The error introduced by low-precision floating-point or fixed-point computations can in general be modeled as white noise if unbiased rounding techniques are used in all arithmetic operations. 
A widely employed rounding technique is \emph{round half to even}, which does not introduce a systematic bias, and is used by default in IEEE 754 floating-point arithmentic~\cite{IEEE2019}. In the following, we assume the usage of such a rounding technique also for fixed-point arithmetic, leading to an only unbiased error within the computed interatomic forces. 

To demonstrate the concept of approximate computing, we introduce white noise to the interatomic forces that are computed while running the MD simulation. In this section, we describe in detail how we introduce the noise to mimic in software the behavior that would be observed when running the MD on the actual FPGA or GPU hardware with reduced numerical precision. We classify the computational errors into two types: fixed-point errors, and floating-point errors. Assuming that $\textbf{F}_{I}$ are the exact and $\textbf{F}_{I}^{N}$ the noisy forces from a MD simulation with low precision on an FPGA for instance, fixed-point errors can by modelled by

\begin{equation}
\textbf{F}_{I}^{N}=
\begin{pmatrix}
\text{F}_{I}^{x}\\
\text{F}_{I}^{y}\\
\text{F}_{I}^{z}\\
\end{pmatrix} +
\begin{pmatrix}
c_{1} \times 10^{-\beta }\\
c_{2} \times 10^{-\beta }\\
c_{3} \times 10^{-\beta }\\
\end{pmatrix}
,
\end{equation}

\noindent whereas floating-point errors are described by

\begin{equation}
\textbf{F}_{I}^{N} =
\begin{pmatrix}
\text{F}_{I}^{x} \times 10^{-\alpha_1}\\
\text{F}_{I}^{y} \times 10^{-\alpha_2}\\
\text{F}_{I}^{z} \times 10^{-\alpha_3}\\
\end{pmatrix} +
\begin{pmatrix}
c_{1} \times 10^{-(\alpha_1+\beta)}\\
c_{2} \times 10^{-(\alpha_2+\beta)}\\
c_{3} \times 10^{-(\alpha_3+\beta)}\\
\end{pmatrix}
.
\end{equation}

\noindent Therein, $c_1$, $c_2$ and $c_3$ are random values chosen in the range [-0.5, 0.5], whereas $\text{F}_{I}^{x}$, $\text{F}_{I}^{y}$ and $\text{F}_{I}^{x}$ are the individual force components of $\textbf{F}_{I}$, respectively. The floating-point scaling factor is denoted as $\alpha$ and the magnitude of the applied noise by \(\beta\).

To rigorously correct the errors introduced by numerical noise we employ a modified Langevin equation. In particular, we model the force as obtained by a low-precision computation on a GPU or FPGA-based accelerator as

\begin{equation} \label{fFPGA}
\textbf{F}_{I}^{N} = \textbf{F}_{I} + \mathbf{\Xi }_{I}^{N},
\end{equation}

\noindent where $\mathbf{\Xi }_{I}^{N}$ is an additive white noise for which

\begin{equation} \label{CrossCorr}
 \left \langle \textbf{F}_{I}\left ( 0 \right ) \mathbf{\Xi } _{I}^{N}\left ( t \right )\right \rangle \cong  0
\end{equation}

\noindent holds. Throughout, $\langle \cdots \rangle$ denotes Boltzmann-weighted ensemble averages as obtained by the partition function $Z=\text{Tr} \exp(-E/k_B T)$, where $E$ is the potential energy, $k_B$ the so-called Boltzmann constant, and $T$ the temperature. Given that $\mathbf{\Xi }_{I}^{N}$ is unbiased, which in our case is true by its very definition, it is nevertheless possible to accurately sample the Boltzmann distribution by means of a Langevin-type equation \cite{Krajewski,Richters,Karhan}, which in its general form reads as

\begin{equation} \label{LangevinEq}
M_{I}\ddot{\textbf{R}}_{I}=\textbf{F}_{I}+\mathbf{\Xi }_{I}^{N}-\gamma _{N}M_{I}\dot{\textbf{R}}_{I},
\end{equation}

\noindent where $\dot{\textbf{R}}_{I}$ are the nuclear coordinates (the dot denotes time derivative), $M_I$ are the nuclear masses and $\gamma _{N}$ is a damping coefficient,
which is chosen to compensate for \(\mathbf{\Xi }_{I}^{N}\). The latter, in order to guarantee an accurate canonical sampling, has to obey
the fluctuation-dissipation theorem

\begin{equation}
\left \langle \mathbf{\Xi }_{I}^{N}\left ( 0 \right ) \mathbf{\Xi }_{I}^{N}\left ( t \right ) \right \rangle \cong  2 \gamma_{N} M_I k_{B} T  \delta \left ( t \right ).
\label{FDT}
\end{equation}

\noindent Substituting Eq.~\ref{fFPGA} into Eq.~\ref{LangevinEq} results in the desired modified Langevin equation

\begin{equation} \label{modLangevin}
M_{I}\ddot{\textbf{R}}_{I} = \textbf{F}_{I}^{N}-\gamma _{N}M_{I}\dot{\textbf{R}}_{I},
\end{equation}

\noindent which will be used throughout the remaining of this paper. This is to say that the noise, as originating from a low-precision computation, can be thought of as the additive white noise of a damping coefficient $\gamma_N$, which satisfies the fluctuation-dissipation theorem of Eq.~\ref{FDT}. The specific value of $\gamma_N$ is determined in such a way so as to generate the correct average temperature, as measured by the equipartition theorem

\begin{equation}
\left\langle \frac{1}{2} M_I \dot{\textbf{R}}_{I} \right\rangle = \frac{3}{2} k_B T.
\label{EquiPartTheorem}
\end{equation}

\section{Computational details}
\label{sec:computational}

\begin{table}
  \caption{Values for \textit{\(\gamma_N^{fix}\)} and \textit{\(\gamma_N^{float}\)} as a function of \textit{\(\beta\)}.}
  \centering
  \label{tab:gamma}
  \begin{tabular}{lrr}
    \textit{\(\beta\)} & \textit{\(\gamma_N^{fix}\)} & \textit{\(\gamma_N^{float}\)} \\
    \hline
    0 &           & 0.00025  \\
    1 & 0.0004    & 0.000005 \\
    2 & 0.000009  & 0.000005 \\
    3 & 0.0000009 &
  \end{tabular}
\end{table}

To demonstrate our approach we have implemented it in the CP2K suite of programs \cite{CP2Ka, CP2Kb}. More precisely, we have conducted MD simulations of liquid Silicon (Si) at 3000~K using the environment-dependent interatomic potential of Bazant et al. \cite{EIP1,EIP2}.
All simulations consisted of 1000 Si atoms in a 3D-periodic cubic box of length 27.155~\AA. Using the algorithm of Ricci and Ciccotti \cite{Ricci}, Eq.~\ref{LangevinEq} was integrated with a discretized timestep of 1.0~fs with $\gamma_N = 0.001~$fs$^{-1}$.

Whereas the latter settings were used to compute our reference data, in total six different cases of fixed-point and floating-point errors were investigated by varying the exponent $\beta$ between 0 (huge noise) and 3 (tiny noise) that is, ranging from $1/1000$ of the physical force up to the same magnitude as the force.
As already alluded to above, the additive white noise is compensated via Eq.~\ref{modLangevin} by 
continously adjusting the friction coefficient $\gamma_N$ using the adaptive Langevin technique of Leimkuhler and coworkers so as to satisfy the equipartition theorem of Eq.~\ref{EquiPartTheorem}~\cite{JonesLeimkuhler2011, Mones2015, LeimkuhlerStoltz2019}. In this method, $\gamma_N$ is reinterpreted as a dynamical variable, defined by a negative feedback loop control law as in the Nos\'e-Hoover scheme~\cite{Nose,Hoover}. The corresponding dynamical equation for $\gamma_N$ reads as

\begin{equation}
  \dot{\gamma}_N= (2K-n k_B T)/\mathcal{Q},
\end{equation}

\noindent where $K$ is the kinetic energy, $n$ is the number of degrees of freedom and $\mathcal{Q}=k_B T \tau^2_{NH}$ is the Nose-Hoover fictitious mass with time constant $\tau_{NH}$. Alternatively, $\gamma_N$ can be estimated by integrating the autocorrelation function of the additive white noise \cite{RZK}.
In Table~\ref{tab:gamma} the resulting values of \textit{\(\gamma_N^{fix}\)} for fixed-point and \textit{\(\gamma_N^{float}\)} for floating-point errors are reported as a function of \textit{\(\beta\)}.

\section{Results and Discussion}
\label{sec:results}
As can be directly deduced from Table~\ref{tab:gamma}, the smaller values of $\gamma_N$ for a given $\beta$ immediately suggest the higher noise resilience when using floating-point as compared to fixed-point numbers.
\begin{figure}
\begin{center}
\includegraphics[width=0.8\textwidth]
{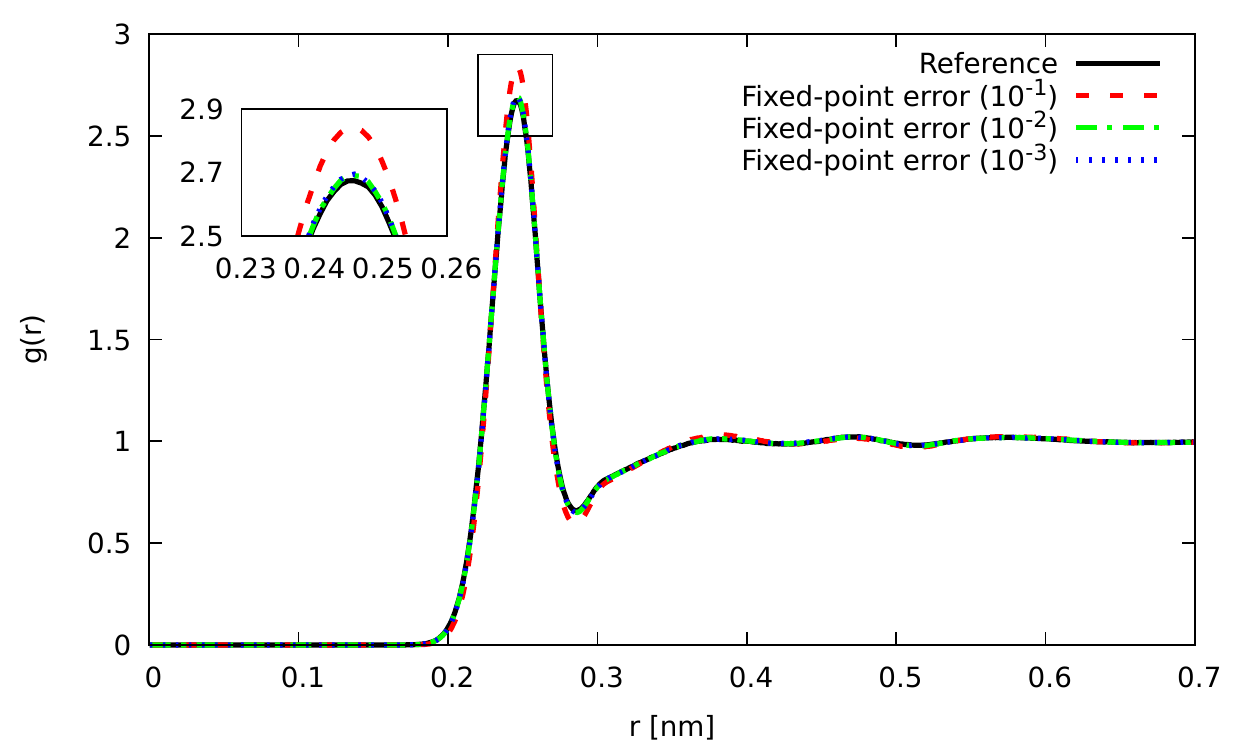}
\end{center}
\caption{\label{Fig1}
Partial pair correlation function for liquid Si at 3000~K with noisy forces introduced by fixed-point errors of magnitude $10^{-3}$ (blue), $10^{-2}$ (green) and $10^{-1}$ (red). For comparison, the results, as obtained by our reference calculations without noise, are shown in black.
} \end{figure}
In Figs.~\ref{Fig1} and \ref{Fig2}, the Si-Si partial pair-correlation function $g(r)$, which describes how the particle-density varies as a function of distance from a reference particle (atoms, molecules, colloids, etc.), as computed using an optimal scheme for orthorombic systems \cite{KAF}, is shown for different values of $\beta$.
As can be seen, for both fixed-point and floating-point errors, the agreement with our reference calculation is nearly perfect up to the highest noise we have investigated. As already anticipated earlier, the usage of floating-point errors is not only able to tolerate higher noise levels, but is also throughout more accurate.
\begin{figure}
\begin{center}
\includegraphics[width=0.8\textwidth]
{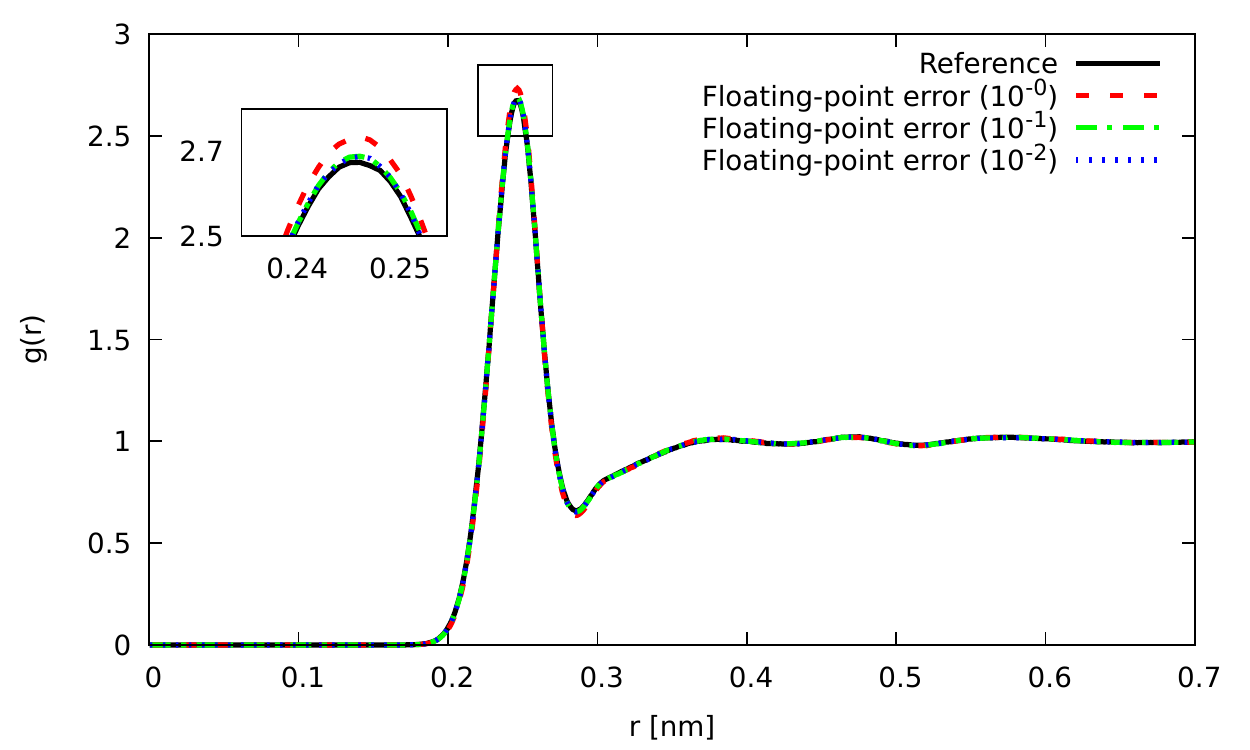}
\end{center}
\caption{\label{Fig2}
Partial pair correlation function for liquid Si at 3000~K with noisy forces introduced by floating-point errors of magnitude $10^{-2}$ (blue), $10^{-1}$ (green) and $10^{-0}$ (red). For comparison, the results, as obtained by our reference calculations without noise, are shown in black.
} \end{figure}

\begin{figure}
\begin{center}
\includegraphics[width=0.8\textwidth]
{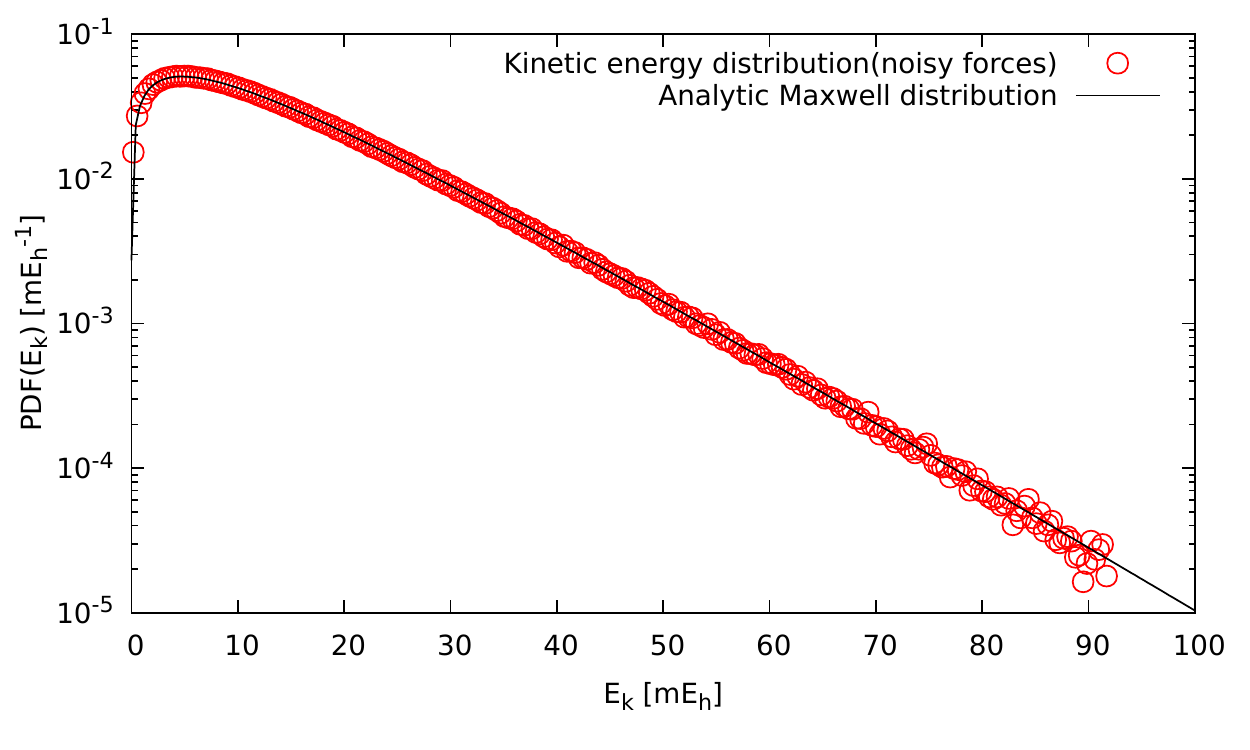}
\end{center}
\caption{\label{Fig3}
Kinetic energy distribution of liquid Si at 3000~K, as obtained by our simulations using noisy forces (circles). For comparison the analytic Maxwell distribution is also shown (line).
} \end{figure}

To verify that the sampling is indeed canonical, in Fig.~\ref{Fig3} the actual kinetic energy distribution as obtained by our simulations using noisy forces is depicted and compared to the analytic Maxwell distribution. It is evident that if sampled long enough, not only the mean value, but also the distribution tails are in excellent agreement with the exact Maxwellian kinetic energy distribution, which demonstrates that we are indeed performing a canonical sampling.
To further assess the accuracy of the present method, we expand the autocorrelation of the noisy forces, i.e.

\begin{subequations}
\begin{eqnarray}
  && \left \langle \textbf{F}_{I}^{N}\left ( 0 \right )\textbf{F}_{I}^{N}\left ( t \right )\right \rangle \\
  &=& \left \langle \left ( \textbf{F}_{I}\left ( 0 \right ) + \mathbf{\Xi } _{I}^{N} \left(0 \right )\right) \left( \textbf{F}_{I}\left ( t \right )+\mathbf{\Xi } _{I}^{N}\left ( t \right )\right) \right \rangle \\
  &=& \left \langle \textbf{F}_{I}\left ( 0 \right ) \textbf{F}_{I}\left ( t \right )\right \rangle + \left \langle \textbf{F}_{I}\left ( 0 \right ) \mathbf{\Xi } _{I}^{N}\left(t \right )\right \rangle \label{AutoCorr} \\
  &+& \left \langle \textbf{F}_{I}\left ( t \right ) \mathbf{\Xi } _{I}^{N}\left(0 \right )\right \rangle + \left \langle \mathbf{\Xi } _{I}^{N}\left(0 \right ) \mathbf{\Xi } _{I}^{N}\left(t \right )\right \rangle.  \nonumber
\end{eqnarray}
\end{subequations}

\begin{figure}
\begin{center}
\includegraphics[width=0.8\textwidth]
{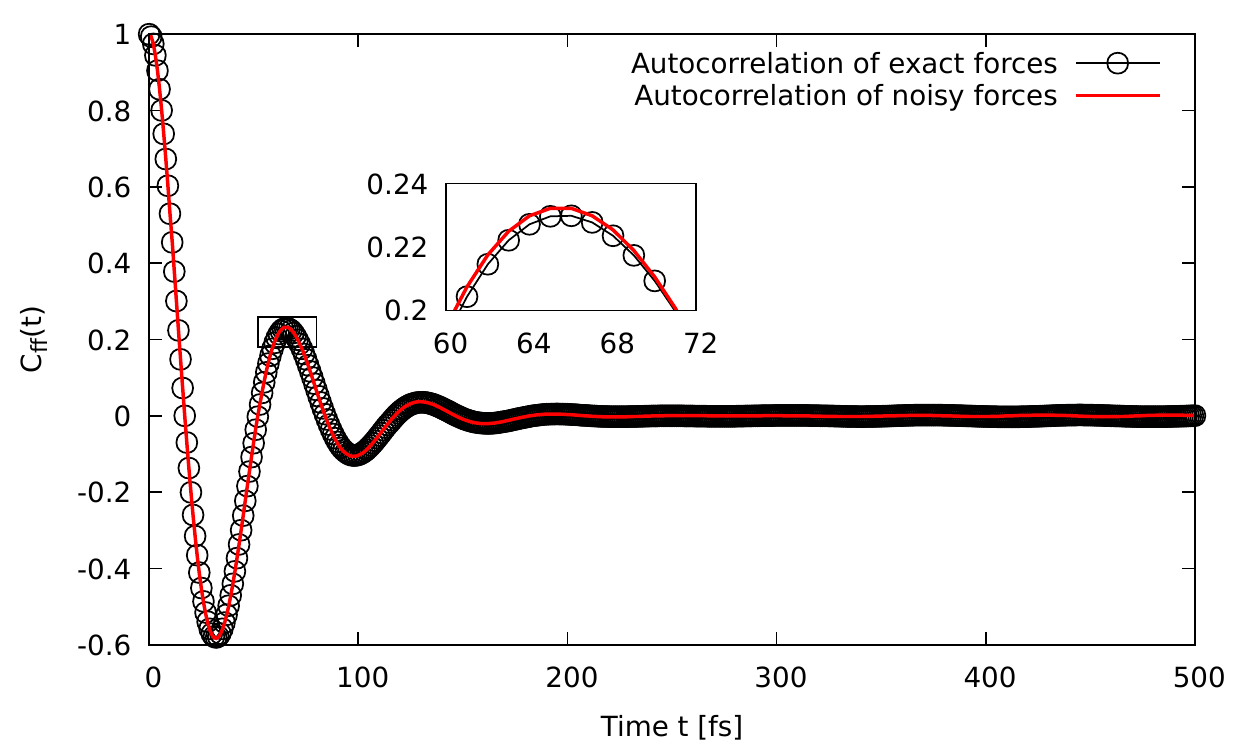}
\end{center}
\caption{\label{Fig4}
The Autocorrelation of the noisy forces \(
\left \langle \textbf{F}_{I}^{N}\left ( 0 \right ) \textbf{F}_{I}^{N}\left ( t \right )\right \rangle \)(line), which are compared to the autocorrelation of the exact forces \( \left \langle \textbf{F}_{I}\left ( 0 \right ) \textbf{F}_{I}\left ( t \right )\right \rangle \)(circles).
} \end{figure}

\noindent Since the cross correlation terms between the exact force and the additive white noise is vanishing due to Eq.~\ref{CrossCorr}, comparing the autocorrelation of the noisy forces $\langle \textbf{F}_{I}^{N}(0)\textbf{F}_{I}^{N}(t)\rangle$ with the autocorrelation of the exact forces $\langle \textbf{F}_{I}(0) \textbf{F}_{I}(t)\rangle$ permits to assess the localization of the last term of Eq.~\ref{AutoCorr}.
The fact that $\langle \textbf{F}_{I}^{N}(0)\textbf{F}_{I}^{N}(t)\rangle$ is essentially identical to $\langle \textbf{F}_{I}(0) \textbf{F}_{I}(t)\rangle$, as can be seen in Fig.~\ref{Fig4}, implies that $\langle \mathbf{\Xi } _{I}^{N}(0) \mathbf{\Xi } _{I}^{N}(t)\rangle$ is very close to a $\delta$-function as required by the fluctuation-dissipation theorem in order to ensure an accurate canonical sampling of the Boltzmann distribution. In other words, from this it follows that our initial assumption underlying Eq.~\ref{modLangevin}, to model the noise due to a low-precision calculation as an additive white noise channel, is justified.

\section{Conclusion}
\label{sec:conclusion}
We conclude by noting that the present method has been recently implemented in the universal force engine i-PI \cite{iPi}, which can be generally applied to all sorts of forces affected by stochastic noise such as those computed by GPUs or other hardware accelerators~\cite{HOOMD, NAMD, OpenMM, HalMD, Lammps, Amber, Gromacs}, and potentially even quantum computing devices \cite{Steane, Knill, Blatt, Chow}. The possibility to apply similar ideas to N-body simulations~\cite{White, Makino} and to combine it with further algorithmic approximations~\cite{LassAC} is to be underlined and will be presented elsewhere.

\funding{
The authors would like to thank the Paderborn Center for Parallel Computing (PC$^2$) for computing time on \textsc{OCuLUS} and FPGA-based supercomputer \textsc{Noctua}. Funding from the Paderborn University's research award for ``Green IT'' is kindly acknowledged. This project has received funding from the European Research Council (ERC) under the European Union's Horizon 2020 research and innovation programme (Grant Agreement No.:~716142) and from the German Research Foundation (DFG) under the project PerficienCC (grant agreement No PL 595/2-1).}

\reftitle{References}



\end{document}